\documentclass[letterpaper]{jpconf}
\usepackage[pdftex]{graphicx}
\usepackage{iopams}
\usepackage[numbers,square,comma,sort&compress]{natbib}
\usepackage{fancyvrb}
\usepackage[colorlinks,bookmarks=false,citecolor=blue,linkcolor=red,urlcolor=blue]{hyperref}

\def\gsim{\mathrel{\rlap{\lower3pt\hbox{\hskip0pt$\sim$}}\raise1pt\hbox{$>$}}}    
\VerbatimFootnotes

\begin{document}
\title{Vectorising the detector geometry to optimize particle transport}

\author{John Apostolakis, Ren\'e Brun, Federico Carminati, Andrei Gheata and Sandro Wenzel} 
\address{European Organization for Nuclear Research (CERN), Geneva, Switzerland}
\ead{sandro.wenzel@cern.ch}

\begin{abstract}
Among the components contributing to particle transport, geometry
navigation is an important consumer of CPU cycles. The tasks performed
to get answers to "basic" queries such as locating a point within a
geometry hierarchy or computing accurately the distance to the next
boundary can become very computing intensive for complex detector
setups. So far, the existing geometry algorithms employ mainly scalar
optimisation strategies (voxelization, caching) to reduce their CPU
consumption. In this paper, we would like to take a different approach and
investigate how geometry navigation can benefit from the vector
instruction set extensions that are one of the primary source of performance
enhancements on current and future hardware.  While on paper, this form
of microparallelism promises increasing performance opportunities,
applying this technology to the highly hierarchical and multiply
branched geometry code is a difficult challenge.  We refer to the
current work done to vectorise an important part of the critical
navigation algorithms in the ROOT geometry library. Starting from a
short critical discussion about the programming model, we present the
current status and first benchmark results of the vectorisation of some
elementary geometry shape algorithms.  On the path towards a full
vector-based geometry navigator, we also investigate the performance
benefits in connecting these elementary functions together to develop
algorithms which are entirely based on the flow of vector-data. To
this end, we discuss core components of a simple vector navigator that is
tested and evaluated on a toy detector setup.
\end{abstract}

\section{Introduction}
The Geant-Vector prototype is one of the initiatives \cite{GeantV2012CHEP, GPUprototypeWhitepaper} that try
to recast High-Energy Physics (HEP) particle transport codes into a
form that allows them to benefit from all performance
dimensions on current and future (commodity) hardware -- in particular
from both multicore-
(multithreading) and micro- parallelism (vectorisation).
By doing so, these projects try to overcome the increasing gap between the performance of existing software and the ideal performance limits coming from advances in computing technology \cite{OpenLabCHEP2012}.

One of the important activities in the context of the Geant-Vector prototype is to factorise the CPU-intensive algorithms contributing to particle transport and to try to reshuffle both geometry and physics code in a vectorisable form.
While first efforts were focused on investigating concurrency issues \cite{GeantV2012CHEP}, the project now also addresses the opportunities offered by vector microparallelism.

The choice to start our investigation on the
geometry component was stimulated by the fact that geometry
calculations traditionally consume a considerable CPU budget in a
typical detector simulation (up to $40-50\%$ of the transport time). It is
our hope that the experience gained within this activity will
help us implement similar strategies in other simulation components,
such as physics processes, at a later stage.

Starting from a short review of vectorisation approaches, we will
report below on the status of vectorising some of the shape navigation algorithms 
in the ROOT geometry library \cite{TGeoDoc}. Building on top of these vectorised algorithms, we will then discuss the results for the higher level task of finding the distance to the next boundary within a volume that contains several ``daughter volumes'' making up a toy detector model. This benchmark allows us to estimate performance benefits according to
different origins (gains due to refactoring, vectorisation, code
locality, etc.), which will be helpful for extrapolating to more complex
algorithms.

\section{\label{sec:vec}Vectorisation basics and programming models}
Vector-units executing \emph{single instructions on multiple
  data} (SIMD) in parallel were introduced in commodity
hardware in the late 90`s in form of extensions to
the x86 architecture and instruction set (for example MMX, SSE, 3DNOW, AVX).
These vector extensions allow the simultaneous application of particular instructions (say an \emph{add} instruction) on multiple data elements 
leading to an increase in throughput compared to scalar (serial) handling of the same data elements.
Current SIMD extensions (AVX, AVX2) can handle simultaneous operations
on 4 doubles (8 floats) at the same time whereas vector units able to
handle 8 doubles are already available on the Intel Xeon Phi and are
announced in the upcoming AVX3 instruction set extension. 

How can we make use of SIMD? The necessary prerequisite is the
availability of (contiguous) data on which the same operations should
be carried out.  In the Geant-Vector prototype, this is being realised
by grouping the particles in the same logical volume (potentially
from different events) into a data-parallel container called
a \emph{basket} \cite{GeantV2012CHEP}.  Basic algorithms can then
process particles within the same basket in a loop-like fashion.  In
this circumstance, modern compilers are in principle able to generate SIMD
vector instructions by autovectorising the relevant loop.  In reality,
our experience is that this currently works only in a few cases
(such as simple array operations in tight loops) and often requires
substantial refactoring of the code with only a few handles that
developers can control. Finally, autovectorisation is usually an
all-or-nothing procedure that, for instance, makes it hard to mix
vector and scalar operations; as an example, a single scalar function call
within a loop will usually prevent autovectorisation.

At the other extreme, complete control can be given to the developer
by directly coding in a vector-oriented way using intrinsics or
assembly instructions. Higher-level alternatives to this are currently
available in the form of (non-standardized) vector types, available in
some compilers or in the form of portable C++ template libraries
\cite{Vc2012,BOOSTSIMD} that encapsulate the low-level details in
template classes with a C++ high-level syntax (operator overloading).
These libraries still require a reformulation of existing original
code but allow, to a very large extent, for portable and maintainable code. 
For example, the Vc library is a free software library that eases explicit vectorisation of C++ code \cite{Vc2012,Vc2012homepage} and which provides for us the
essential advantage to write code in which
SIMD vector operations are easily mixed with scalar code. Overall we
have gained a good experience with Vc and the results reported below
have all been obtained by making use of  Vc version 0.73.

However, it should be noted that once we have formulated our
algorithms in terms of vector operations, a move to alternative vector-oriented programming models/languages seems to be straightforward. Good
experience has already been obtained from \emph{directly}
translating Vc code into Intel CilkPlus \cite{IntelCilkPlus} array
notation. Equally, it should be straightforward to extend the vector-oriented
code to different platforms such as accelerators (for instance GPUs). Ultimately, we have to
see how a single code base can be practically maintained for all
targeted architectures. We are currently starting to
investigate how the evolving standards OpenMP/OpenACC
\cite{OPENMP4,OPENACC} can help in this respect.

\section{Towards a vector-oriented geometry navigator}
The main CPU usage related to geometry during particle transport 
is due to the execution of navigation algorithms. The geometry navigator calls several elementary algorithms for each
particle/track to calculate parameters such as geometrical steps, safe
distances to (volume) boundaries and to determine the containing volume (localization) of particles within a complex detector geometry.
\begin{figure}
\centering
\includegraphics[width=0.5\textwidth]{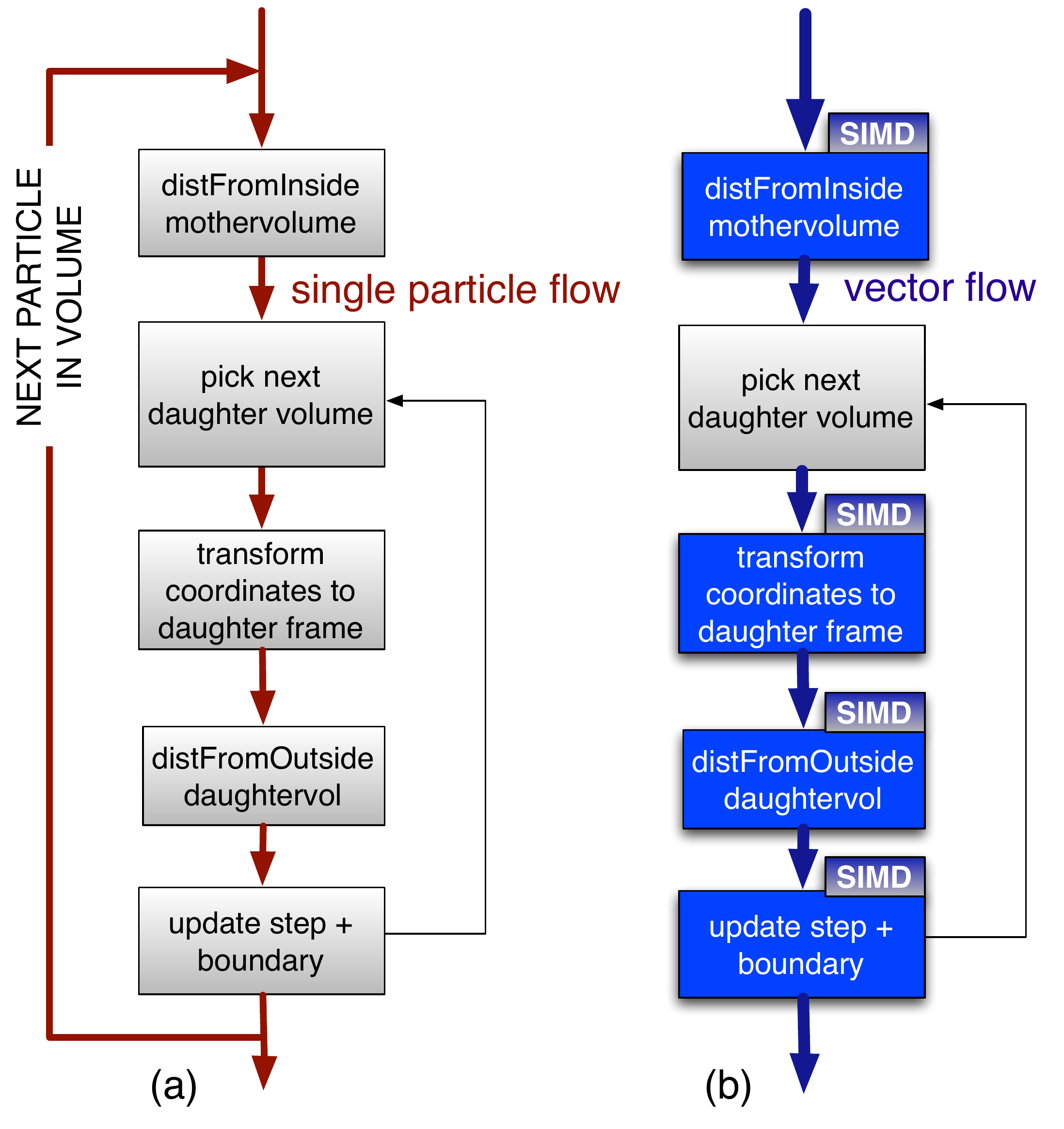}
\caption{\label{fig:navigator}Representation of the core part of a geometry navigator addressing the problem of 
obtaining the distance to the next hit boundary of particles in a certain detector volume (which itself contains further daughter volumes).
(a) Schema of a sequential algorithm versus (b) an algorithm
based on flow of vectors. The two essential components are the functions  \texttt{distFromInside} (mothervolume) and \texttt{distFromOutside} (daughtervolume), which are used to determine the distance of a particle to the boundary of its current detector element as well as the distance to other volumes contained within this element along its line of flight.}
\end{figure}
In order to make efficient use of the SIMD capabilities and the
availability of vector data in form of baskets \cite{GeantV2012CHEP,
  GeantV2013CHEP}, the Geant-Vector prototype has to
provide a geometry navigator that is able to process vectors of
particles instead of operating on single particles/tracks in a
sequential mode. Figure \ref{fig:navigator} sketches this concept for the core functionalities of the navigator by calculating the
geometrical step length to the next boundary along the particle line of flight and the corresponding hit
volume. This algorithm is missing the final localization of
  particles in the detector geometry which has been left out for the
  first investigation. Note also that this is a rather simple form of
  a navigator without voxelization techniques. In
Figure~\ref{fig:navigator}a, the algorithm is based on processing
single particles through a series of elementary algorithms while in
Figure~\ref{fig:navigator}b the same algorithm is presented based on
a flow of vector data. Essentially, going from the scalar version to
the SIMD optimized vector version of the algorithm requires two steps,
namely a refactoring of the API to enable passing vectors across algorithms,
followed by the adaptation of each elementary algorithm to profit as much as possible from SIMD
instructions. Note that both steps
individually contribute to a performance gain over the scalar version.  As will be shown in
Section~\ref{sec:complexalgos}, the first step decreases the number of
function calls, reduces the number of memory moves and improves code
locality, while the second step increases throughput via
microparallelism.

\subsection{\label{sec:simplealgos}Elementary geometry algorithms}
\begin{figure}
\centering
\includegraphics[width=0.65\textwidth]{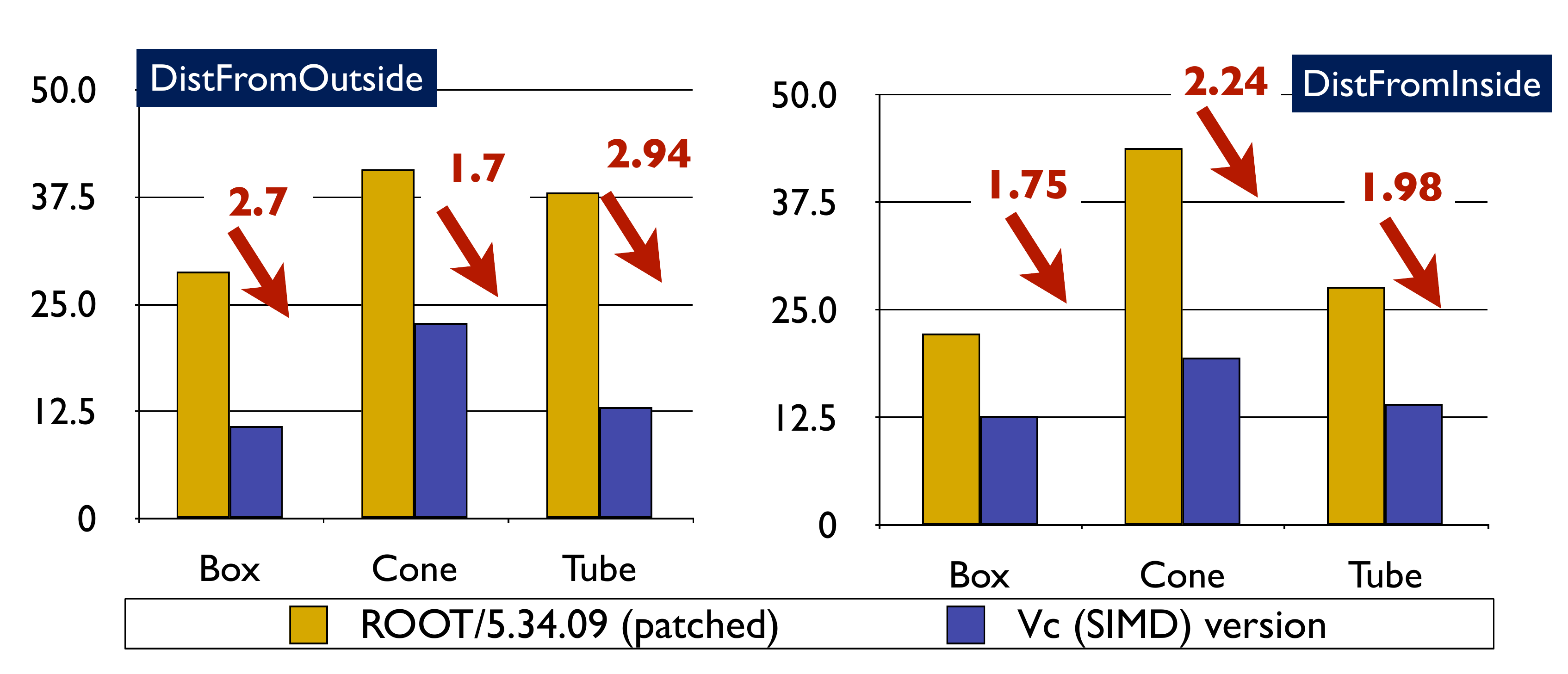}
\includegraphics[width=0.34\textwidth]{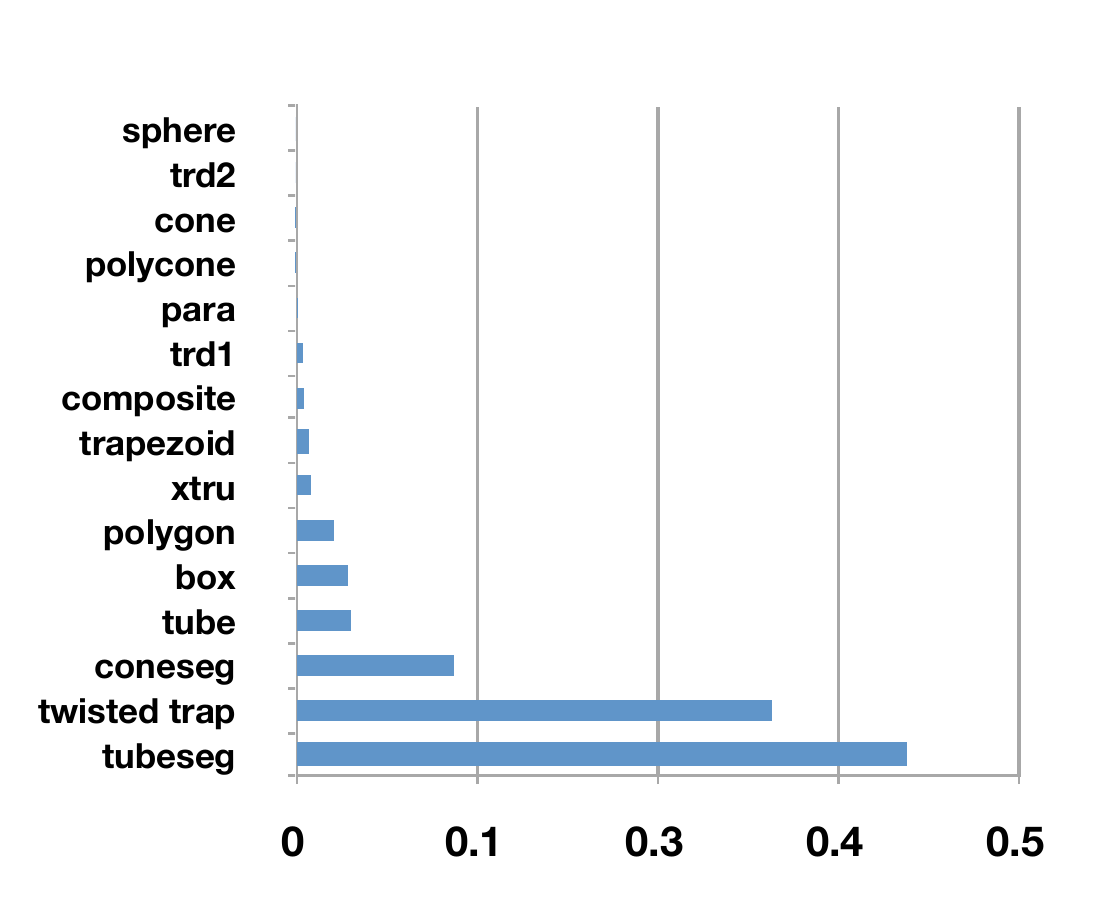}
\caption{\label{fig:simplealgs}Speedups obtained from vectorizing
  simple algorithms for the box, cone and tube shapes from the ROOT shape library. The functions presented are \texttt{distFromOutside} and
 \texttt{distFromInside}, i.e. the distance to the shape boundary from a point outside and from inside the shape respectively. We also show a simple estimate of the
  relative CPU budget (in percentage on the x-axis) for various shapes based on counting physical
  shapes in detectors of $33$ existing HEP experiments and taking into
  account the scalar runtime cost for the function \texttt{distFromOutside}.}
\end{figure}
In order to progress towards a full vector geometry navigator, we have to
provide the basic geometry algorithms (single blocks in
Figure~\ref{fig:navigator}a) in a recast and optimized form. Among the most important basic algorithms are 
the calculations of the particle distances to
the solids/shapes composing the detector.
We hence started our investigation of the potential of SIMD
instructions at the level of the shape methods, notably on the functions
that calculate the distance of the particles, along the line of flight, to the inside (entering) of a shape or its 
distance to get outside (leaving), as well as on functions that calculate the minimum distance to any boundary (safety) and functions performing inside/outside checks.
The current vectorisation work is based on the existing code in the ROOT geometry library \cite{TGeoDoc}, but the extension to future standard libraries such as the Unified Solids \cite{USolidsCHEP2012} is foreseen.

The first shape tackled was the box since it is one of the most simple yet
important geometrical forms (see Figure~\ref{fig:simplealgs} for
an averaged estimate of the importance of various shapes).  The
simplicity of the box gave us a good playground to quickly assess the
various programming models and memory layouts.  Considering the
difficulties of trying to get the code to autovectorise, versus the
relative ease of programming (and compiler independence) with a
library like Vc, we then opted for the second choice for the purpose
of this first performance evaluation.  At the time of writing, several
of the simple shapes, such as boxes, cones, tubes (including their
segmented forms), have been successfully ported to Vc
code. Figure~\ref{fig:simplealgs} gives an overview of the speedup
achieved so far for the most important methods.  These benchmarks were
run on an Intel Ivy Bridge machine (\verb+Intel(R) Core(TM) i7-3770 CPU @ 3.40GH+, SLC6
  \verb+2.6.32-358.18.1.el6.x86_64+, \verb+gcc-4.7+, \verb+Vc-0.73+,
  \verb+ROOT-5.34.09+) with AVX instruction set and results are
reported for 1024 particles in a
basket. These speedups, measured at between 1.7 and 3, are matching our expectations and provide a first confirmation that we are
moving into the right direction.

Refactoring and optimization work on more complicated shapes such as
polycones (sequence of connected cone and tube segments) has
started. First positive results for polycones with few $z$ segments have been already
obtained, while for polycones with many $z$ segments the availability of optimized
sequential algorithms (with voxelization and table lookup techniques)
makes it hard to better expose the parallelism in a way that can be exploited
by SIMD-capable hardware because different
particles will in general follow slightly different code paths. 
It will be one of the future challenges to adapt these special sequential optimisations in complicated shapes with microparallelism.
However, note that we will anyway gain from the re-factored
interface of those shapes alone by exploiting better code locality
and less function calls.

Besides the shape algorithms, we have successfully
SIMD-optimised various other elementary algorithms needed by a
vector-oriented geometry navigator, such as coordinate transformations
needed for coordinate frame conversions or min/max algorithms.

\subsection{\label{sec:complexalgos}From elementary to complex vector algorithms}
Building on top of the vector-enabled basic components, we have made a
first implementation of the vector-oriented geometry navigator as
shown in Figure~\ref{fig:navigator}b. This simple approach allows
for a more realistic evaluation of the performance gains coming from
the combined usage of vector-optimised algorithms used in the vector
prototype.
\begin{figure}
\centering
\begin{minipage}{0.45\textwidth}
\includegraphics[width=\textwidth]{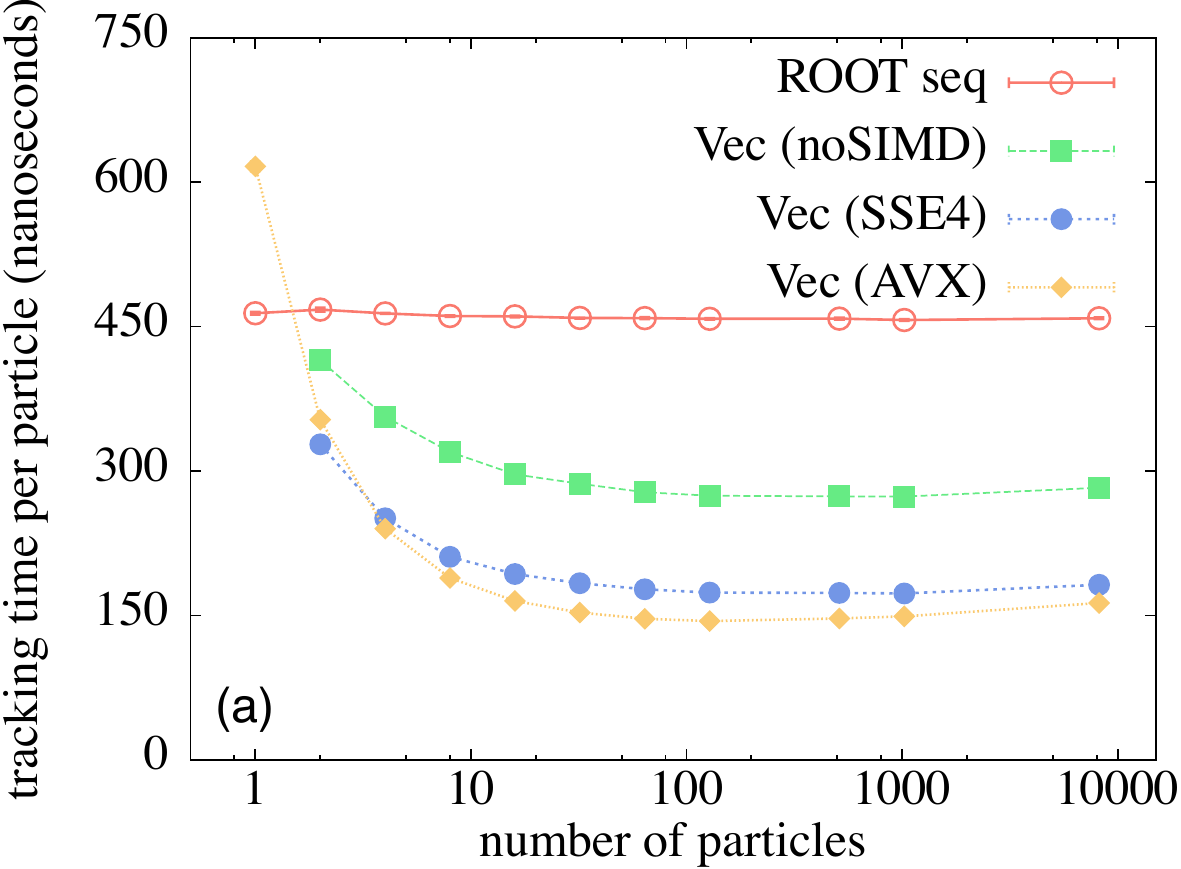}
\end{minipage}
\makebox[1cm]{}
\begin{minipage}{0.45\textwidth}
\includegraphics[width=\textwidth]{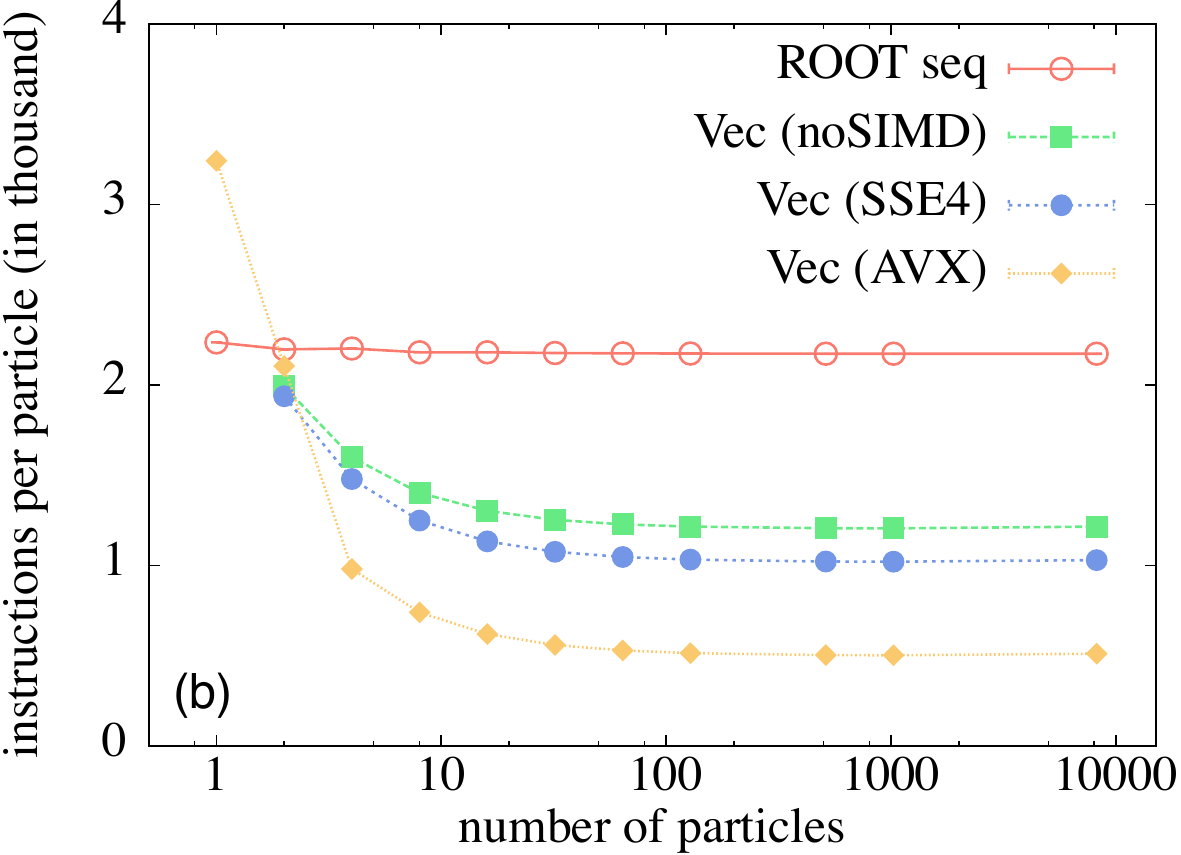}
\end{minipage}
\begin{minipage}{0.46\textwidth}
\includegraphics[width=\textwidth]{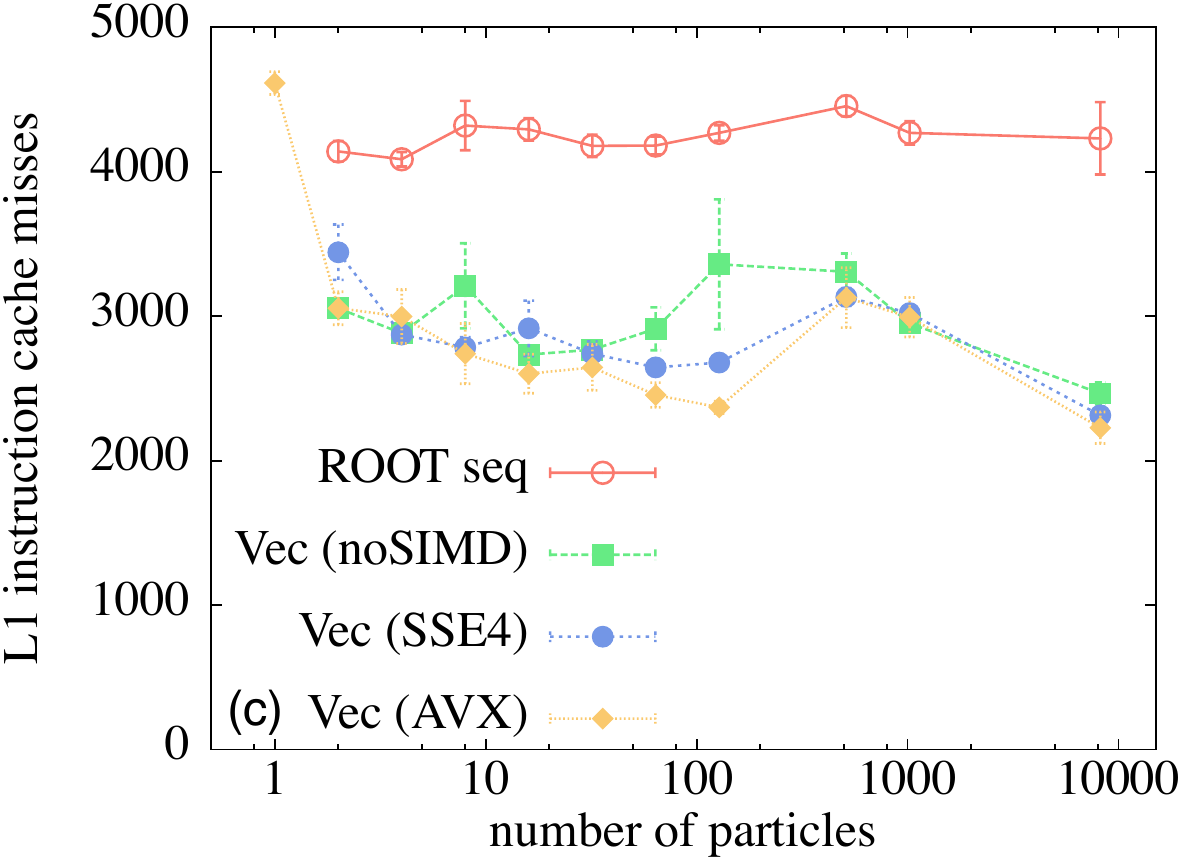}
\end{minipage}
\makebox[0.5cm]{}
\begin{minipage}{0.48\textwidth}
\caption{\label{fig:resultsnav}Results from the benchmark comparing the scalar algorithm (ROOT seq) with a vector-oriented algorithm using various degrees of usage of SIMD instructions [VEC(noSIMD), VEC(SSE4) and VEC(AVX)]. (a) Comparison of the runtime per particle showing a speedup factor of roughly $3$ comparing the original version to the AVX code. (b) Comparison of the number of instructions executed per particle. (c) Comparison of level-$1$ instruction cache misses (arbitrary scale). }
\end{minipage}
\end{figure}

To this end, we compare the performance of the sequential algorithm
(Figure~\ref{fig:navigator}a) using the standard scalar approach from
the existing ROOT package with the newly implemented version. To
estimate a best-case scenario, a small toy detector setup was made out of already optimised shapes and which should serve as a first
(standard) benchmark. The toy detector consists of a tubular mother
volume containing two other tubes (beampipe, shield),
four boxes representing detector plates and two cones (as endcaps).
The intent was to have a non-trivial setup for the simple algorithm presented here. For more complex setups we would have to combine vectorisation with scalar optimization techniques, such as voxelization, which has to be addressed in a future step.

We filled the exclusive part
of the mother volume with a large pool of random particle positions
and directions. To run the benchmark, we pick $N$ consecutive
particles in memory starting from a random position in this pool and process them with both
the sequential and vector algorithms to obtain the distances and next
hit boundaries for all of them. We repeat this process $P$
times and in each benchmark we keep the product of $N\times P$
constant to give the same amount of work to each
benchmark run.

Besides comparing the runtime between the scalar and vectorised
navigators, we also looked into different metrics given by the
hardware performance counters of the CPU. We could directly measure
and compare the code locality in terms of L1 instruction cache misses,
branch misses and data cache misses. For this we interface with the
perfmon library \cite{perfmon} with which we are able to read out the
counters right before and after the specific code
section. Additionally, details of the actual instructions executed in
the relevant code sections are obtained using a custom binary
instrumentation tool using the Intel Pin API \cite{pintool}.

A couple of key results from this study are shown in
Figure~\ref{fig:resultsnav} where data is included for the original
scalar and sequential algorithm, for the refactored
algorithm based on vector flow but without SIMD optimization,
and for the vector algorithms with SIMD optimisations (labeled in the Figure as ROOTseq, Vec(noSIMD), VEC(SSE4$|$AVX) respectively). Our main result is that we
are currently able to speed-up the example algorithm of
Figure~\ref{fig:navigator} by a factor of $\gsim 3$ and that
considerable gains are even seen for rather small number of
particles in a basket. The speedup originates from various
contributions as shown in the plots: just by refactoring into a
vector interface [version VEC(noSIMD)], a performance gain with a factor $\approx
1.5$ is seen in terms of the runtime. The SIMD instructions, using instruction sets SSE4 or AVX, then give the actual gains from
microparallelism.
We can track the origin of the gains by analysing the dynamic
instruction mix actually executed in the benchmark. Some important
numbers obtained from this analysis are summarised in
Table~\ref{tab:instructions}. Going from a scalar to vector interface allows to reduce the number of
function call instructions accompanied by a massive reduction in simple memory moves (such
as those used to save registers on the stack). When introducing SIMD
optimisations with Vc, the overall number of instructions further
shrinks and the CPU vector unit is used to a much higher degree.
\begin{table}
\centering
\caption{\label{tab:instructions}Statistics
  for instructions executed in the algorithm of
  Figure~\ref{fig:navigator}. The first line of numbers shows the relative reduction of the total instruction count (ALL) with respect to the sequential algorithm. The numbers in the second block show the fractions for simple memory moves (MOV), call
  instructions (CALL), SIMD instructions (ALL SIMD) as well as
  arithmetic SIMD instructions (ARITHM SIMD) relative to the total
  number of instructions within each column. These numbers are obtained
  for $16$ particles and a comparison is done between the four
  algorithmic versions mentioned in the text.}

\begin{tabular}{l*{200}llll}
\br
Instruction (type) & ROOT seq & VEC(noSIMD) & VEC(SSE4) & VEC(AVX) \\
\mr
ALL         & $1$      & $0.6$    & $0.52$     & $0.29$    \\
\mr
MOV         & $0.296$  & $0.116$   & $0.132$   & $0.163$  \\
CALL        & $0.036$  & $0.0023$  & $0.0026$  & $0.0048$ \\
ALL SIMD    & $0.043$  & $0.188$   & $0.641$   & $0.57$   \\
ARITHM SIMD & $0.023$  & $0.039$   & $0.289$   & $0.30$   \\
\br
\end{tabular}
\end{table}
Using the hardware performance counters, we have also confirmed that
the number of instruction cache misses is considerable reduced due to
better code locality when using the vectorised interfaces (Figure \ref{fig:resultsnav}c). We expect
this effect to become even more important with more complex
algorithms.

\section{\label{sec:conclusions}Summary and Outlook}
Focusing on the geometry component in particle transport codes, we
described the status of our vectorisation effort within the
Geant-Vector prototype.  On the basis of an explicit vector-oriented
programming model with a high level C++ template library (Vc), we reported on significant performance
improvements for important distance methods of simple shape classes in the ROOT
geometry library. 
These SIMD improvements together with a new vector API, allowing to pass vectors of data across the basic algorithmic components, add up to 
a total performance gain of the order of $300\%$ for the example of a simple vector-oriented navigation algorithm in a toy-geometry.

There are a multitude of challenges to be tackled in the future: the simple navigation
algorithm above has to be extended to be a full geometry navigator, we
have to SIMD optimize more complex geometrical shapes and most
importantly we have to figure out how more complicated algorithms
(voxelization) can make use of SIMD instructions. We have also to
extend and validate our findings to other hardware, such as GPUs.

\ack
We thank Marilena Bandieramonte, Raman Sehgal, Juan Valles Martin
for contributions to the Vc coding. We also would like to thank Pere Mato, Vincenzo Innocente (PH-SFT, CERN), Laurent Duhem (Intel), Andrzej
Nowak (Openlab, CERN) and Matthias Kretz (University of Frankfurt) for very stimulating discussions.

\bibliographystyle{iopart-num}
\bibliography{literature}

\providecommand{\newblock}{}
\begin{thebibliography}{10}
\expandafter\ifx\csname url\endcsname\relax
  \def\url#1{{\tt #1}}\fi
\expandafter\ifx\csname urlprefix\endcsname\relax\def\urlprefix{URL }\fi
\providecommand{\eprint}[2][]{\url{#2}}

\bibitem{GeantV2012CHEP}
Apostolakis J, Brun R, Carminati F and Gheata A 2012 {\em Journal of Physics:
  Conference Series\/} {\bf 396} 022014
  \urlprefix\url{http://stacks.iop.org/1742-6596/396/i=2/a=022014}

\bibitem{GPUprototypeWhitepaper}
{Canal} P, {Elvira} D, {Hatcher} R, {Jun} S~Y and {Mrenna} S 2013 {\em ArXiv
  e-prints\/} (\textit{Preprint} \eprint{1307.7452})

\bibitem{OpenLabCHEP2012}
Jarp S, Lazzaro A and Nowak A 2012 {\em Journal of Physics: Conference
  Series\/} {\bf 396} 052058
  \urlprefix\url{http://stacks.iop.org/1742-6596/396/i=5/a=052058}

\bibitem{TGeoDoc}
Gheata A The root geometry
  \urlprefix\url{ftp://root.cern.ch/root/doc/18Geometry.pdf}

\bibitem{Vc2012}
Kretz M and Lindenstruth V 2012 {\em Software: Practice and Experience\/} {\bf
  42} 1409 \urlprefix\url{http://dx.doi.org/10.1002/spe.1149}

\bibitem{BOOSTSIMD}
Esterie P, Gaunard M, Falcou J, Laprest{\'e} J~T and Rozoy B 2012 {\em PACT\/}
  ed Yew P~C, Cho S, DeRose L and Lilja D~J (ACM) p 431 ISBN 978-1-4503-1182-3
  \urlprefix\url{http://doi.acm.org/10.1145/2370816.2370881}

\bibitem{Vc2012homepage}
The Vc~homepage
  \urlprefix\url{http://code.compeng.uni-frankfurt.de/projects/vc}

\bibitem{IntelCilkPlus}
The Intel CilkPlus~homepage \urlprefix\url{https://www.cilkplus.org}

\bibitem{OPENMP4}
OpenMP-Consortium \urlprefix\url{http://openmp.org/wp/openmp-specifications/}

\bibitem{OPENACC}
OpenACC-Consortium 2013 \urlprefix\url{http://www.openacc-standard.org}

\bibitem{GeantV2013CHEP}
Apostolakis J, Brun R, Carminati F, Gheata A and Wenzel S 2013 {\em Journal of
  Physics: Conference Series (CHEP2013)\/}

\bibitem{USolidsCHEP2012}
Gayer M, Apostolakis J, Cosmo G, Gheata A, Guyader J~M and Nikitina T 2012 {\em
  Journal of Physics: Conference Series\/} {\bf 396} 052035
  \urlprefix\url{http://stacks.iop.org/1742-6596/396/i=5/a=052035}

\bibitem{perfmon}
The~perfmon/libpfm homepage \urlprefix\url{http://perfmon2.sourceforge.net/}

\bibitem{pintool}
The Intel Pin~homepage \urlprefix\url{http://www.pintool.org}

\end{thebibliography}
\end{document}